\documentclass[preprint,aps,prl,superscriptaddress,amssymb,showpacs]{revtex4}
\usepackage[dvips,final]{graphicx}
\begin{document}
\title{Ferromagnetic- and superconducting-like behavior of the electrical
resistance of inhomogeneous graphite flake} %%
\author{J. Barzola-Quiquia}
\affiliation{Division of Superconductivity and Magnetism, Institut
f\"ur Experimentelle Physik II, Universit\"{a}t Leipzig,
Linn\'{e}stra{\ss}e 5, D-04103 Leipzig, Germany}
\author{P. Esquinazi}\email{esquin@physik.uni-leipzig.de}
\affiliation{Division of Superconductivity and Magnetism, Institut
f\"ur Experimentelle Physik II, Universit\"{a}t Leipzig,
Linn\'{e}stra{\ss}e 5, D-04103 Leipzig, Germany}

\begin{abstract}
We have measured the magnetic field and temperature dependence of
the resistivity of several micrometers long and heterogeneously
 thick  graphite sample.   The magnetoresistance results for
 fields applied nearly parallel to the graphene planes show both a
granular superconducting behavior as well as the existence of
magnetic order in the sample.
\end{abstract} \pacs{74.10.+v,74.45.+c,75.50.Pp} \maketitle

%\section{Introduction}
The possible existence of high-temperature superconductivity in
highly oriented pyrolytic graphite (HOPG) has been speculated 10
years ago  from resistivity  and SQUID measurements
\cite{kopejltp07}. For fields parallel to the c-axis of the HOPG
samples and after subtraction of the huge diamagnetic background, the
magnetization showed superconducting-like, sample-dependent
hysteresis loops \cite{kopejltp07}. Strikingly, SQUID measurements
indicated also the presence of high-temperature magnetic order in
graphite. Both behaviors appeared to depend on the sample and its
thermal treatment. The intrinsic origin of the ferromagnetic signals
in HOPG in virgin  as well as in irradiated HOPG found convincing
evidence  through  SQUID, x-ray magnetic circular dichroism (XMCD)
and anisotropic magnetic resistance (AMR) measurements, see
Ref.~\onlinecite{jems08} and Refs. therein. There is consent that
this phenomenon is related to defects in the graphite structure,
including non-magnetic ad-atoms as hydrogen
\cite{kusakabe03,duplock04,yaz07,pis07,zha07}.

Regarding the existence or not of superconductivity in graphite
several studies provided some but still not conclusive support for
its existence, see Ref.~\onlinecite{kopejltp07} and Refs.~therein.
Theoretical work that deals with possible superconducting states in
graphite as well as in graphene has been published in recent years.
For example, $p$-type superconductivity has been predicted to occur
in inhomogeneous regions of the graphite structure \cite{gon01} or
$d-$wave high-$T_c$ superconductivity in graphite based on resonance
valence bonds \cite{doni07}. Following the BCS approach in two
dimensions (with anisotropy) critical temperatures $T_c \sim 60~$K
have been estimated if the density of conduction electrons per
graphene plane is $n \sim 10^{14}~$cm$^{-2}$, a density that might be
induced by defects and/or hydrogen ad-atoms \cite{garbcs09}.
Predictions for superconductivity in graphene were also published
\cite{uch07,kop08} supporting the idea that $ n > 10^{13}~$cm$^{-2}$
in order to reach $T_c  > 1~$K. In contrast to the basically 3D
superconductivity in intercalated graphitic compounds, see e.g.
Ref.~\onlinecite{csa05}, it is speculated that $T_c$ in the graphene
layers of graphite may be much larger because of the role of the
high-energy phonons of the 2D graphite structure itself
\cite{garbcs09}.

Recently published high-resolution measurements of the
magnetoresistance (MR) of micrometer small and several tens of
nanometers thick graphite samples suggest the existence of
inhomogeneous, granular superconductivity with critical temperature
$T_c \gtrsim 25~$K \cite{esq08}. The claim of granular
superconductivity in graphite is based on three main observations:
(a) the existence of anomalous hysteresis loops of resistance versus
magnetic field applied parallel to the $c-$axis  that indicate the
existence of Josephson-coupled superconducting grains, (b) the
observation of oscillatory behavior in the magnetoresistance vs.
field that can be related to Andreev reflections of Copper pairs at
the interfaces of superconducting and semiconducting regions, similar
to a phenomenon observed recently in conventional
superconducting-normal junctions \cite{rud08}, and (c) the phenomena
observed in (a) and (b) vanish at $T > T_c$, where $T_c$ denotes the
temperature at which the resistance $R(T)$ reaches a maximum, i.e.
$R(T< T_c)$ shows metallic behavior with an anomalous large
magnetoresistance. The observed phenomena support the view that HOPG
is a system with non-percolative superconducting domains immersed in
a semiconducting-like matrix \cite{esq08}.

Recent experimental study suggests that the internal interfaces
between $\sim 50~$nm thick (in the $c-$axis direction) crystalline
graphite regions  running parallel to the graphene planes might be
the regions where superconductivity is located \cite{bar08}. These
interfaces may have enough carrier density to trigger granular
superconductivity with different critical temperatures, keeping the
quasi-two dimensional behavior. This phenomenon is actually not new
but already observed  at the interfaces between Bi crystals where
superconductivity up to $\simeq 20~$K was found, although pure Bi is
not superconducting \cite{mun08}.

The aim of this work is to check whether transport measurements could
provide further evidence for a granular superconducting-like
behavior. For this purpose we need to locate the voltage electrodes
nearer those internal interfaces. Simultaneously, we search for
ferromagnetic-like behavior in the MR, which should exist in virgin
HOPG samples according to SQUID results for fields applied parallel
to the graphene planes \cite{jems08}. Our experimental work
demonstrates that a heterogeneously thick and several micrometers
long graphite flake  shows superconducting as well as
ferromagnetic-like behaviors in the MR. The prevailing behavior
depends on the measured region on the sample.

%\section{Experimental details}

The sample we show in this study was obtained from a HOPG sample with
$0.4^\circ$ rocking curve width. The impurity levels of metallic
elements are below $5~\mu$g/g with the exception of V ($16~\mu$g/g).
The preparation of the MG samples is simple and based on exfoliation
of pieces of the HOPG sample.  We have used  ac current $I$ of
amplitude $\lesssim 1~\mu$A. Current dependent behavior has been
observed at $I \gtrsim 3~\mu$A but because  self heating effects can
play a role, we will not discuss this in this work. Note also that
the current does not flow only through the interfaces and therefore
the interpretation of current dependence in the MR is not simple.

\begin{figure}[]
%\vspace{0.5cm}
\begin{center}
\includegraphics[width=1.0\columnwidth]{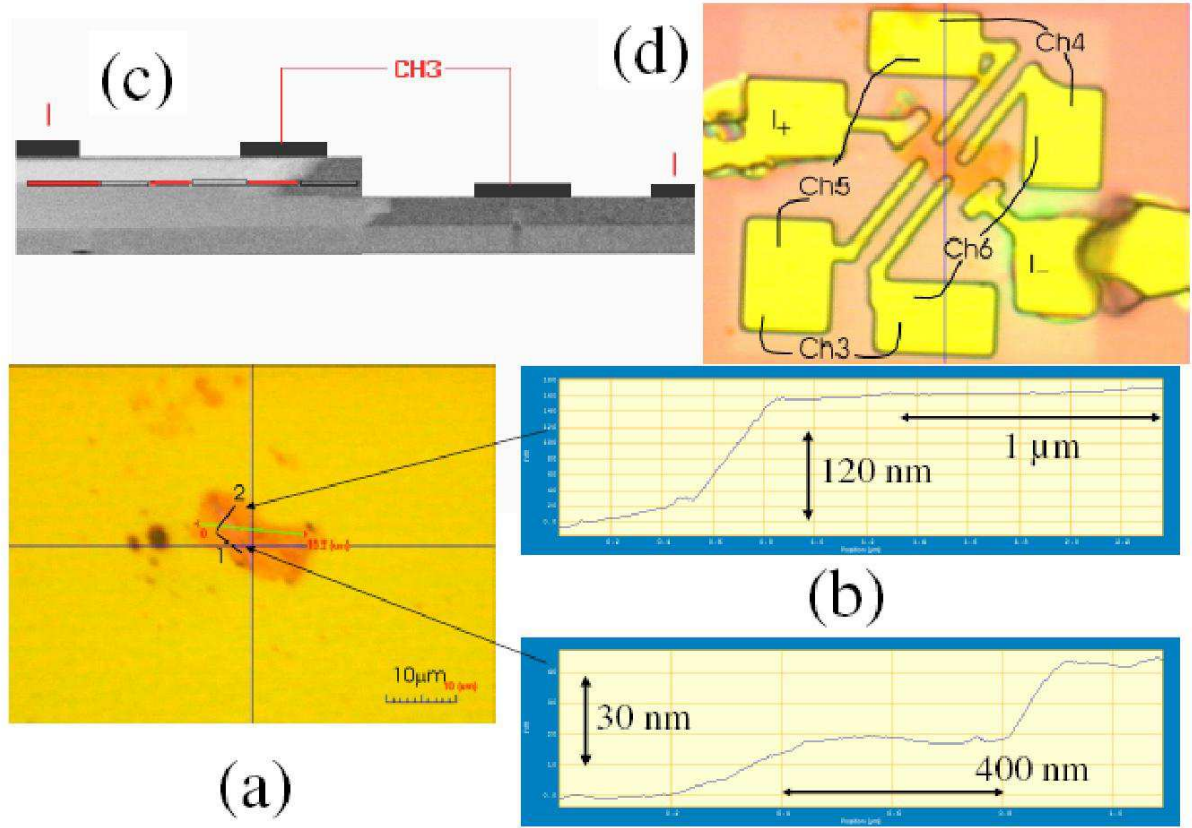}
\caption[]{(a) Optical picture of the MG sample. The lines 1 and 2
give the scan directions of the AFM line scans shown at (b). (c)
Electron transmission microscope picture of a 200 nm thick lamella
obtained from the same bulk HOPG sample as the MG sample. The
$c-$axis of the graphite structure runs normal to the interfaces. The
thickness of the single crystalline regions (defined by the
homogeneous gray color regions) is between 20~nm to $\sim 70$~nm. At
the top of this picture we draw schematically the positions of the
CH3 and current electrodes. In case superconducting regions exist
between the CH3 electrodes, e.g at the red lines depicted in the
picture, separated by semiconducting regions (gray lines), we expect
to see some evidence in the voltage drop. (d) Optical microscope
photo of the sample with its electrode distribution.} \label{photo}
\end{center}
\end{figure}

Figure~\ref{photo}(a) shows an optical microscope picture of the
measured $\sim 200~$nm thick sample without electrodes (sample S2A).
The pictures in Fig.~\ref{photo}(b) show the AFM signals measured at
two different regions (see line scans in (a)). In the lower sample
region the AFM scan shows several steps of $\sim 30~$nm height each.
Taking into account the internal structure of HOPG, see the TEM
picture in Fig.~\ref{photo}(c) and also Ref.~\onlinecite{bar08}, some
of the steps are related to  the interfaces between crystalline
regions than run inside the sample parallel to the AFM plateaus. From
the lower to the upper part of the sample (path 2 in (a)) there is an
increase of thickness of $\sim 100~$nm, see upper AFM scan in
Fig.~\ref{photo}(b). Therefore, we decided to put two pairs of
voltage electrodes at the lower and upper parts of the sample,
labeled CH3 and CH4 and  the electrodes for current at the two
extrema $I_{+}, I_{-}$, see Fig.~\ref{photo}(d). In
Fig.~\ref{photo}(c) we have drawn schematically the positions of the
CH3 electrodes as well as some of the expected granular
superconducting regions (red straight lines parallel to the graphene
planes and the interface region) separated by semiconducting regions
(gray lines) at one of the interfaces. The measurements presented
below were done on the longitudinal channels CH3 and CH4. The other
channels CH5 and CH6 show a mixture of usual Hall,  planar Hall
effect features and longitudinal resistance behavior, as observed in
CH3 and CH4  when the field is applied nearly parallel to the sample
surface, and will not be further discussed here.

%\section{Transport properties}

\begin{figure}[]
%\vspace{0.5cm}
\begin{center}
\includegraphics[width=0.95\columnwidth]{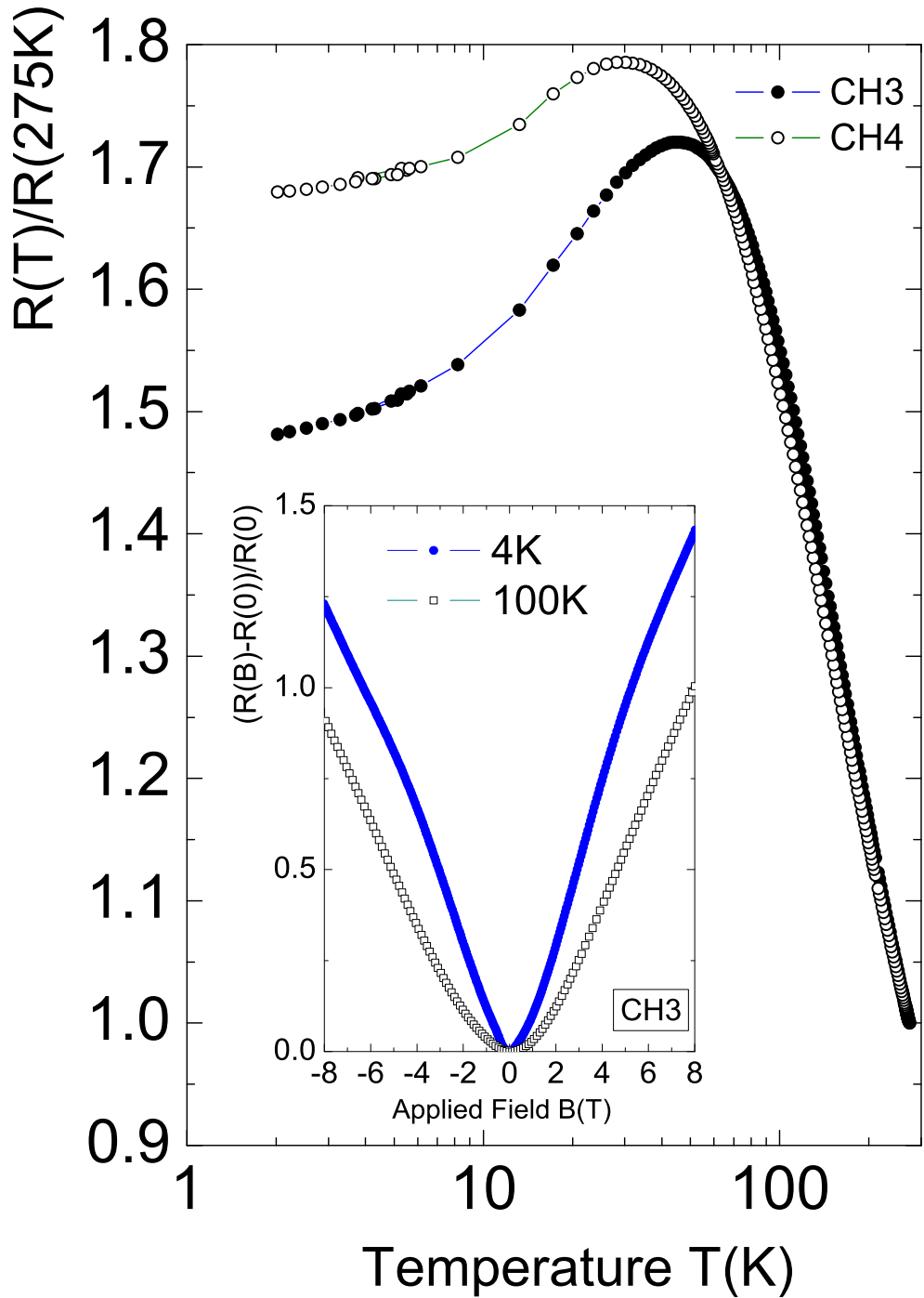}
\caption[]{Temperature dependence at zero applied field of the
normalized resistance for the two longitudinal channels CH3 and
CH4. The inset shows the magnetoresistance $MR$ at two
temperatures for CH3 and fields applied normal to the graphene
planes of the sample.} \label{Tdep}
\end{center}
\end{figure}

Figure~\ref{Tdep} shows the temperature dependence of the
normalized resistance $R(T)$  measured at CH3 and CH4. This
dependence is typical for MG samples of this thickness
\cite{esq08,bar08}, i.e. $R(T)$ shows a semiconducting behavior
with a sample and position dependent resistance maximum. The
semiconducting behavior of $R(T)$ is basically intrinsic of
graphite; it is affected by lattice defects and in particular by
the interfaces between crystalline regions \cite{bar08}. In the
inset of the same figure we show  MR $= R(B)-R(0)/R(0)$ at two
temperatures for CH3 and for fields applied normal to the graphene
planes of the sample. Although the zero-field resistances at 4~K
and 100~K are equal,  the MR's differ each other in disagreement
with Kohler's rule. This
 indicates that there is an extra
electronic mechanism that increases by $\sim 25\%$ the MR(8T) at
4~K with respect to 100~K.

Taking into account previous results \cite{esq08,bar08} we
interpret the decrease of $R$ below $\sim 50$K for CH3 as due to
weakly coupled superconducting regions. Decreasing temperature the
phases of the order parameters of individual superconducting
regions become correlated decreasing the total resistance. Within
this hypothesis the whole behavior of $R$ depends on the single
grains condensation energy and the strength of the intergrain
Josephson coupling energy. Global superconductivity is established
via Josephson tunneling if there is sufficiently long-range phase
coherence. If the superconducting regions are localized at the
boundaries or interphases between crystalline graphite regions,
see Fig.~\ref{photo}(c), even in the case a long range macroscopic
phase coherence  would exist at the interfaces, it is clear that
due to the $c-$axis resistance  a zero-resistance state can never
be measured between the electrodes of CH3.

If our assumption of granular superconductivity below $\sim 50~$K
for the region measured by CH3 is correct, the MR results
 shown in Fig.~\ref{Tdep} suggest that the contribution to the MR
 due to granular superconductivity is overwhelmed by a large background, whatever
its origin. Because our aim is to search for a more obvious hint
for superconductivity in the MR, it should be clear that we need
to decrease the MR background  substantially. This is possible
reducing the field component normal to the graphene layers
\cite{kempa03} in such an amount that the applied field will still
be able to influence the order parameter of the superconducting
grains and/or the coupling between them.

\begin{figure}[]
%\vspace{0.5cm}
\begin{center}
\includegraphics[width=1.\columnwidth]{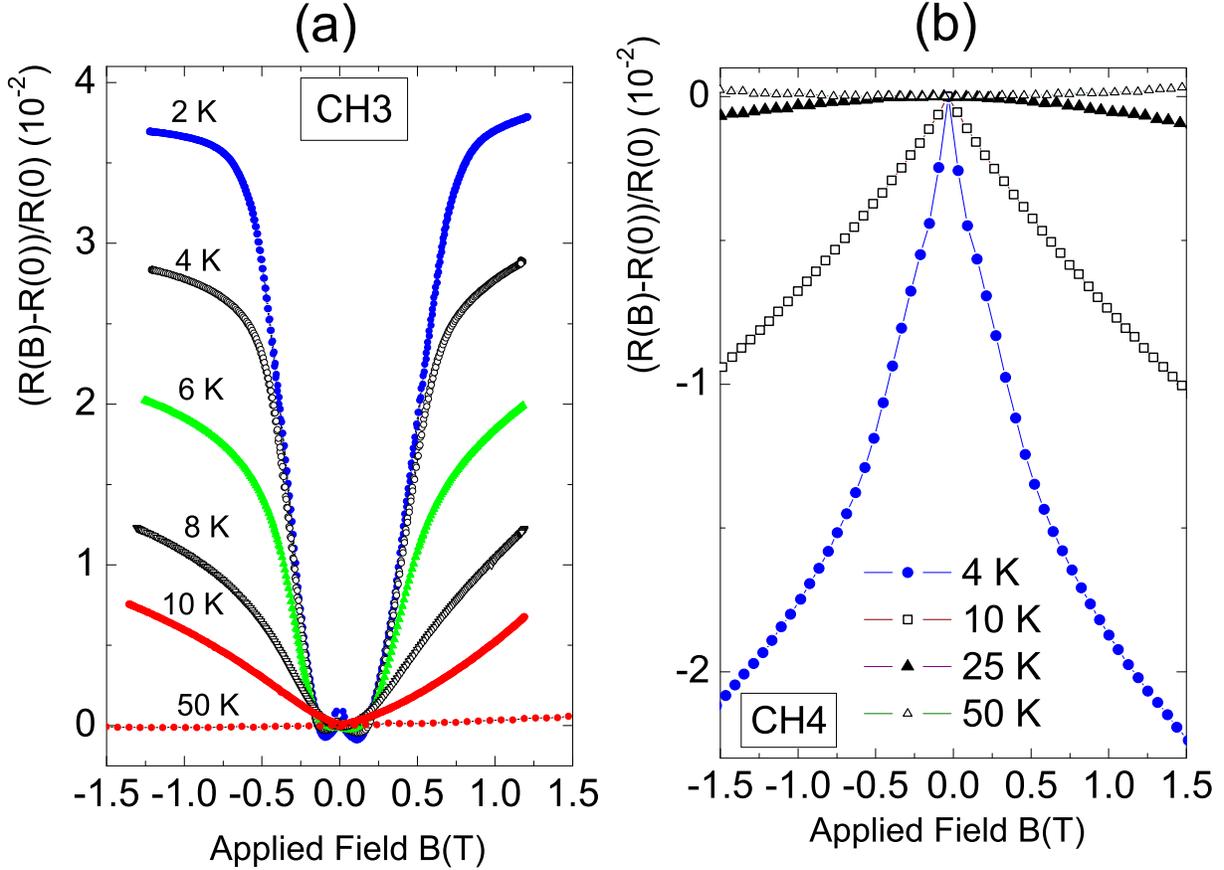}
\caption[]{(a) Magnetoresistance measured at CH3 vs. magnetic
field applied parallel to the graphene planes within $\pm 3^\circ$
at different temperatures. (b) The same but for CH4. The current
was applied at an angle of 55$^\circ$  with respect to the
magnetic field. The saturation of the MR is observed up to 8~T,
the highest field applied.} \label{mr}
\end{center}
\end{figure}

Figure~\ref{mr} shows the MR measured at CH3 (a) and CH4 (b) at
different temperatures below 50~K. The magnetic field was applied
parallel to the main area of the sample within  $\pm 3^\circ$. The MR
of CH3 shows an unique behavior not yet reported so far for graphite.
At low $T$ the MR increases abruptly at $B \gtrsim 0.25~$T and
saturates at $B \gtrsim 0.75~$T. According to recently done
theoretical estimates, the carrier density as well as the effective
mass of the carriers in HOPG can have a strong influence on its $T_c$
\cite{garbcs09}. Therefore, we expect a distribution of $T_c$ at the
interfaces. We interpret the broadening of the transition at higher
$T$ as related partially to the expected distribution of $T_c$, to
the increasing phase fluctuations with $T$ but also to a Hall
component, which is the origin of the field asymmetry  in the MR
clearer observed at higher $T$, see Fig.~\ref{mr}(a).

At low enough $T$  the  overall behavior of the MR measured at CH3
resembles that found in different granular superconductors. The
 negative and small MR observed at low fields and
the change to a positive MR,  see Fig.~\ref{mr}(a), are features
strikingly similar to that observed in, e.g. fractal Pb films
\cite{xio97,wan07}, Al nanowires \cite{san89}, or Sn stripes close
to the global $T_c$ \cite{kad78}. Several interpretations about
the origin of the negative MR have been published, as e.g.
non-equilibrium charge imbalance at normal-superconducting
interfaces produced by localized phase-slip centers
\cite{kad78,san89}, quenching of the pair-breaking effect by the
applied field on Kondo impurities localized at the interfaces
\cite{xio97}, increase of the local critical current (determined
by the interference between regions of different superfluid
densities) with field \cite{xio97} or to the coexistence of two
superconducting phases \cite{wan07}. In our case, however, the
negative MR at low fields and temperatures has a different origin,
as we demonstrate below.

\begin{figure}[]
%\vspace{0.5cm}
\begin{center}
\includegraphics[width=1.1\columnwidth]{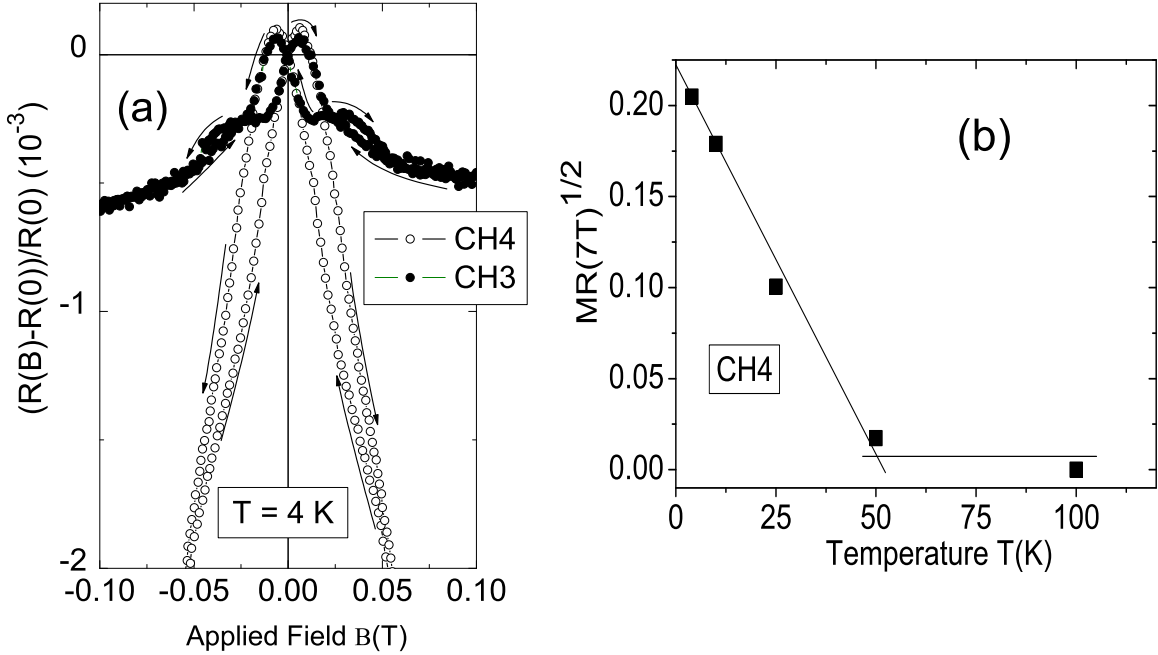}
\caption[]{(a) Magnetoresistance of CH3 and CH4 at 4~K for magnetic
fields applied parallel to the sample graphene planes and at an angle
of $55^\circ$ from the applied current direction. (b) Temperature
dependence of the square root of the magnetoresistance at saturation
($B = 7~$T) that is proportional to the ferromagnetic moment at
saturation obtained for CH4. The linear behavior agrees with that
observed in virgin and irradiated graphite samples and it is
compatible with 2D magnetic order \protect\cite{jems08,barzola2}.}
\label{hys}
\end{center}
\end{figure}

The MR of CH4, measured simultaneously with that of CH3, shows a
qualitatively different behavior, see Fig.~\ref{mr}(b). This is
compatible with those observed in many ferromagnetic materials. This
negative MR is accompanied by a hysteresis at low fields, see
Fig.~\ref{hys}, that gets smaller the higher the temperature. This
fact reveals the existence of magnetic order similar to the one
reported recently for irradiated HOPG sample \cite{bar08tms,jems08}.
Note that at low fields the negative MR is practically identical for
both channels, see Fig.~\ref{hys}(a), and that this vanishes for $T
\rightarrow 50~$K for CH4. We can estimate the change of
ferromagnetic magnetization at saturation $M_s$ assuming that $M_s
\propto $~MR$^{1/2}$ at high enough fields. Using the data obtained
at 7~T for the MR at CH4, we obtain a linear decrease of $M_s$ with
$T$ with an effective critical Curie temperature $T_C \sim 50~$K, see
Fig.~\ref{hys}(b). This linear decrease with $T$ of $M_s(T)$ agrees
with the quasi two-dimensional magnetic order in proton-
\cite{barzola2} as well as carbon-bombarded HOPG samples
\cite{xia08}. We note, however, that this Curie temperature is an
effective one since the electrodes measure in parallel  both
ferromagnetic as well as non-ferromagnetic regions that could have
lower resistivity at high $T$ and therefore affect the measured
voltage.  We may conclude therefore that in the sample region
measured by CH4 the ferromagnetic signal overwhelms.

%\section{Conclusion}
In conclusion we have measured the temperature and magnetic field
dependence of the  resistivity of an inhomogeneous, mesoscopic HOPG
sample. The behavior depends on the sample region. In particular in
sample regions where the electrodes  pick up the response of
interfaces between crystalline regions inside the sample, the
behavior of the MR resembles that of granular superconductors. In
agreement with SQUID measurements done in virgin HOPG samples for
fields normal to the $c$-axis, our transport measurements reveal a
ferromagnetic behavior, which is the origin of the negative MR.

\acknowledgements This work has been possible with the support of the
DFG under DFG ES 86/16-1. The authors gratefully acknowledge
discussions with N. Garc\'ia.

\bibliographystyle{apsrev}
%\bibliography{D:/DATA/hopg/magnetic_carbon}
%\bibliography{D:/data/HOPG/magnetic_carbon}

\end{document}